\documentclass[prb,twocolumn,showpacs,preprintnumbers,amsmath,amssymb]{revtex4}

\usepackage{graphicx}
\usepackage{dcolumn}
\usepackage{bm}
\begin{document}


\title{Transport criticality in triangular lattice Hubbard model}

\author{Toshihiro Sato}
\author{Kazumasa Hattori}
\author{Hirokazu Tsunetsugu}

\affiliation{
Institute for Solid State Physics, University of Tokyo, 5-1-5 Kashiwanoha, Kashiwa, Chiba 277-8581, Japan}

\date{\today}

\begin{abstract}
We study electric transport near the Mott metal-insulator transition.
Optical conductivity of the half-filled Hubbard model on a triangular lattice is
calculated based on a cellular dynamical mean field theory including vertex corrections inside the cluster.
By investigating the spectrum at low frequencies, we find that a Drude peak on the metallic side smoothly connects to an ``ingap" peak on the insulating side.
The optical weight of these peaks exhibits a critical behavior with power-law near the Mott critical end point,
$|D-D^*|\propto|U-U^*|^{1/\delta}$.
We find that the critical exponent $1/\delta$ differs from the exponents in the thermodynamics. 

\end{abstract}

\maketitle

Mott transition is one of the central topics in strongly correlated electronic systems. 
This occurs when Coulomb repulsion predominates over electric kinetic energy.
Mott transition takes place inside the paramagnetic region and magnetic frustration plays an important role
to realize it.
A classic three-dimensional example is Cr-doped V$_2$O$_3$,\cite{MT-exp-1} and quasi-two-dimensional $\kappa$-type organic compounds have been intensively studied recently.\cite{MI-tri-1}
Their phase diagram in the parameter space of temperature $T$ and (chemical) pressure $P$ has insulating and metallic phases
which are separated by a line of first-order Mott transition, and this phase boundary terminates at an end point.
This is similar to the liquid-gas transition in classical liquids~\cite{LC-tri-1}
and critical behaviors are expected for various properties around the Mott critical end point.
In recent works, critical behaviors in thermodynamic quantities have been confirmed near the Mott critical end point
and the criticality has been discussed actively.\cite{MTC-dcc-DMFT-1,MI-crit-1}
However, studies on nonequilibrium or transport properties are limited, although the singularity is most prominent
in electric conductivity.\cite{MT-exp-1}
Putative criticality in electric transport is the main issue of this paper.
We will demonstrate its existence at the Mott critical point and analyze its singularity by numerical approach.

Recently, two experimental studies have reported a critical behavior of dc-conductivity $\sigma_{0}$.
Limelette~{\it et al.}~performed a scaling analysis for $\sigma_{0}$ of (V$_{\rm 1-x}$Cr$_{\rm x}$)$_2$O$_3$ near the Mott critical end point,
which indicates the three-dimensional Ising universal class~(UC).\cite{MTC-exp-V}
However, using the similar analysis, Kagawa~{\it et al.}~found that the critical behaviors in the quasi-two-dimensional anisotropic triangular lattice compound
$\kappa$-(ET)$_{2}$Cu[N(CN)$_{2}$]Cl are not classified as any conventional UC's.\cite{MTC-exp-kapp}
There are several proposals to explain the latter.
Imada~{\it et al.} suggested that the unconventional UC is due to a marginal quantum critical point in the two-dimensional system.\cite{MTC-kapptheo-1}
Another theory by Papanikolaou~{\it et al.} proposed that the different exponents come from the fact that conductivity corresponds to energy density of the
Ising model in addiition to magnetization.\cite{MTC-kapptheo-2}

\begin{figure}
\centering
\centerline{\includegraphics[height=2in]{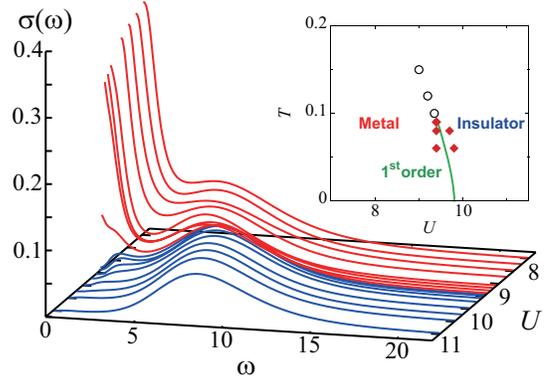}}
\vspace{-0.1cm}
\caption{(Color online) Optical conductivity $\sigma(\omega)$ for various $U$ at $T=0.10$ near the critical point. 
Inset is the $U$-$T$ phase diagram.
Diamond (circle) represents the boundaries of the coexisting phase (crossover region).
The line of the first order transition is a guide for eyes.
}  
\vspace{-0.5cm}
\label{fig:tau0}
\end{figure}

Optical conductivity, $\sigma(\omega)$, also shows peculiar behaviors in the $\kappa$-type organic family,
depending on geometrical frustration in each material.
Strongly frustrated Mott insulator $\kappa$-(ET)$_{2}$Cu$_{2}$(CN)$_{3}$ exhibits a spin liquid behavior, and
its optical conductivity does not show a clear gap but a smooth decay toward zero frequency.\cite{OC-1}
In contrast, $\kappa$-(ET)$_{2}$Cu[N(CN)$_{2}$]Cl is less frustrated due to a distorted triangular structure, and its $\sigma(\omega)$
exhibits a clear gap.\cite{OC-2,OC-3}

In this paper, we numerically study the ``Mott criticality" in transport properties of a strongly correlated electronic system on a geometrically frustrated lattice.
In particular, we focus on optical conductivity near the Mott critical end point and examine its behaviors.

The model to study is the one-band Hubbard Hamiltonian on an isotropic triangular lattice at half filling,
\begin{eqnarray}
H=-t\sum_{\langle i,j \rangle,\sigma}c_{i\sigma}^\dagger c_{j\sigma}+U\sum_{i}n_{i\uparrow}n_{i\downarrow}-\mu\sum_{i,\sigma}
c_{i\sigma}^\dagger c_{i\sigma},
\label{eq:H-1}
\end{eqnarray}
where $t$ is the nearest-neighbor hopping amplitude, $U$ is the on-site Coulomb repulsion and $\mu$ is the chemical potential to tune electron density.
$c_{i\sigma}$ is an electron annihilation operator at site $i$ with spin $\sigma$
and $n_{i\sigma}=c_{i\sigma}^\dagger c_{i\sigma}$.
To calculate optical conductivity, we need single- and two-electron Green's functions.
We compute them by the cellular dynamical mean field theory (CDMFT)\cite{CDMFT} 
to take into account both strong electronic correlations and geometrical frustration. 
We map Hamiltonian (\ref{eq:H-1}) onto a three-site cluster model coupled to an effective medium determined self-consistently.
The cluster Green's functions are computed by using the continuous-time Quantum Monte Carlo (CTQMC) method based on the strong coupling expansion.\cite{CTQMC}

Optical conductivity $\sigma(\omega)$ is calculated from the current-current correlation function $\chi_{\mathbf q}^J(\omega)$ with frequency $\omega$ and wave vector $\mathbf q \rightarrow 0$ 
based on the Kubo formula.
In the single-site DMFT approach, $\chi_{\mathbf q}^J(\omega)$ is usually calculated by convoluting the single-electron Green's function.\cite{OC-DMFT}
To take into account correlation effects, we proceed further beyond the standard formulation and include vertex corrections\cite{VC} inside the cluster,
which is a big challenge in numerical computations.
Our study with CDMFT is one of the first achievements for conductivity and the only preceding work studied a square lattice system
and employed the dynamical cluster approximation.\cite{OC-DCA}
We will explain some details later and first report results.
In what follows, we normalize $\sigma(\omega)$ by the unit of $e^2/\hbar$, where $e$ is the elementary charge and $\hbar$ is the reduced Planck's constant and $U$ and temperature $T$ are in units of $t$.

We start with summarizing the variation of $\sigma(\omega)$ near the Mott critical end point.
The inset of Fig.~\ref{fig:tau0} shows the phase diagram determined by our calculation of double occupancy in the parameter space of $U$ and $T$.\cite{phase-1}
A line of first-order Mott transition separates the metallic phase from the insulating phase,
and this line terminates at the end point, $U^* \sim 9.4$ and $T^* \sim 0.1$.
The main panel shows the variation of $\sigma(\omega)$ with increasing $U$ at $T=0.10$ fixed,
which is nearly $T^*$, and we have checked there is no hysteresis.
For $U \leq 9.4$, $\sigma(\omega)$ shows a Drude peak at low $\omega$, indication of the metallic state,
and an incoherent broad peak around $\omega \sim U$.
For $U \geq 9.5$, the Drude peak disappears quickly and its absence is a characteristic of insulator.
We also calculated $\sigma(\omega)$ at various $T$'s.
At the lower $T=0.09$, $\sigma(\omega)$ shows a jump and hysteresis, corresponding to the first-order transition.
At the higher $T=0.15$, $\sigma(\omega)$ shows a smooth crossover from metal to insulator.
We will investigate expected singularity of $\sigma(\omega)$, and analyze its dependence on $U$ at $T=0.10$ fixed.

\begin{figure}
\centering
\centerline{\includegraphics[height=3.6in]{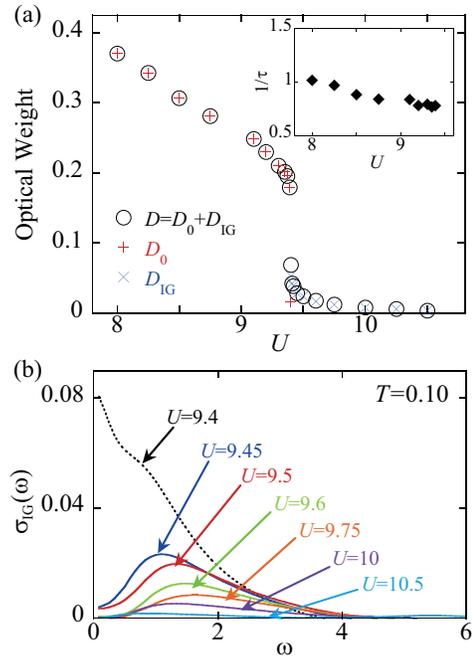}}
\caption{(Color online) (a) Dependence on $U$ of the weights of Drude peak $D_{\rm 0}$, an ingap peak $D_{\rm IG}$, and
total weight $D=D_{\rm 0}+D_{\rm IG}$.
Inset is $U$-dependence of transport scattering rate $1/\tau$ on the metallic side.  
(b) Ingap peak $\sigma_{\rm IG}(\omega)$ on the insulating side for $U \ge 9.4$.
An ingap peak exists at $\omega \sim 1$ and a small Drude peak coexists at $U=9.4$.
}  
\vspace{-0.5cm}
\label{fig:tau4}
\end{figure}

We first discuss the metallic side.
We fit the low energy peak by a simple Drude formula $\sigma(\omega)={\rm Re}[D_{\rm 0}/(-i\omega+1/\tau)]$
and analyze the $U$-dependence of $D_{\rm 0}$ and $1/\tau$.
$D_{\rm 0}$ is a Drude weight, which is inversely proportional to the effective mass $m^*$, and $1/\tau$ is the transport scattering rate.
One may expect that $1/\tau$ diverges with approaching $U^*$.
Instead, we find that $1/\tau$ is almost constant and $D_{\rm 0}$ drops drastically near the Mott transition point, implying the increase of $m^*$ as shown in Fig.~\ref{fig:tau4}~(a).
These results are consistent with an extended Drude analysis including $\omega$-dependence in $m^*$ and $1/\tau$.

On the insulating side, in addition to a broad peak around $\omega \sim U$, $\sigma(\omega)$ has a small peak at a lower $\omega$.
The former persists from the metallic side and comes from the excitations to the Hubbard band.
Analysis reveals that its peak position shifts from the strong coupling limit $\omega \sim U$ by a finite amount $\sim -3$,
and this is due to the motion of doublon-holon pairs.
This Hubbard ``gap" in $\sigma(\omega)$ is not so clear and the incoherent peak is well fitted by a simple
Gaussian form, particularly for $\omega$ below the peak position.
We subtract the Hubbard part from the total $\sigma(\omega)$ and extract the low-$\omega$ part, $\sigma_{\rm IG}(\omega)$.
The results are shown in Fig.~\ref{fig:tau4}~(b).
They show a peak below $\omega \sim 2$, and with approaching $U^*$, the peak position shifts toward lower $\omega$
and the intensity grows quite drastically.
We will call this structure an $\it ingap$ (IG) peak.
The IG peak comes from low-energy peaks with small intensity in the density of states near the Mott transition point.

\begin{figure}
\centering
\centerline{\includegraphics[height=2.25in]{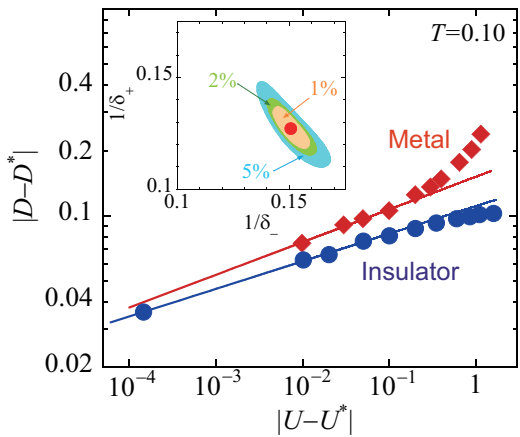}}
\caption{(Color online) Log-log plot for the scaling analysis of the optical weight $D$.
Scaling functions determined by fitting for 4 data points on each side are shown by straight lines.
Inset shows the variation of the exponents with relaxing the fitting error by the amount specified.
}  
\label{fig:tau3}
\vspace{-0.5cm}
\end{figure}

Let us start a detailed analysis of $\sigma(\omega)$ near $U^*$ and examine whether it exhibits a singularity.
Since $\sigma(\omega)$ is not derived from free energy, protocol of its analysis is not standardized like thermodynamic criticality and we try several quantities as a scaling variable. 
The simplest choice is dc-conductivity $\sigma_{0}=\sigma(0)$ and it was used in experimental studies.\cite{MTC-exp-V,MTC-exp-kapp}
However, it turns out that this is not convenient on the insulating side, as we will explain later.
$\sigma_{0}$ is also very sensitive to impurity scatterings, which perturb intrinsic behavior.
An alternative choice of scaling variable is the Drude weight $D_{\rm 0}$ on the metallic side and this is more robust against impurity scatterings.
On the insulating side, however, $D_{\rm 0}=0$, but there exists an IG peak at low-$\omega$, which evolves into a Drude peak on the metallic side.
Thus, it is natural to use its weight $D_{\rm IG}=(2/\pi) \int_{0}^{\infty} d\omega \sigma_{\rm IG}(\omega)$ as a counterpart of $D_{\rm 0}$.

Figure~\ref{fig:tau4} (a) shows the $U$-dependence of $D=D_{\rm 0}+D_{\rm IG}$.
$D$ changes continuously with $U$ but exhibits a singularity; a divergent slope near $U=9.4$.
Following the idea of thermodynamic criticality, we fit the curve with a power-law scaling function,
\begin{eqnarray}
|D-D^*|=A_{\pm}|U-U^*|^{1/\delta_{\pm}},
\label{eq:scal-1}
\end{eqnarray}
where the index $-(+)$ denotes the metallic (insulating) side.
Figure~\ref{fig:tau3} shows the result of scaling analysis using 4 points on each side around the critical point.
The fitting is successful on both sides and $D$ does exhibit the critical behavior.
The critical point is determined simultaneously in this fitting $(U^*, D^*)=(9.3999\pm10^{-4}, 0.11\pm0.01)$.
The error is estimated by relaxing the error minimization in the fitting by 5 \%.

The most important result is the critical exponents, $(1/\delta_{-}, 1/\delta_{+})=(0.15\pm0.02, 0.13\pm0.02)$.
While $1/\delta_{+}=1/\delta_{-}$ as expected within the accuracy, this value is unexpected.
See the inset in Fig.~\ref{fig:tau3}.
In the analysis of experimental data,\cite{MTC-exp-V,MTC-exp-kapp} the fundamental assumption is that deviation in pressure $\Delta P$ from the critical point and
dc-conductivity $\Delta \sigma_{0}$ should be interpreted as magnetic field $h$ and magnetization $m$ in the Ising ferromagnet.
Therefore, the exponent obtained from conductivity is believed to coincide with the one relating $m \propto |h|^{1/\delta}$ at the critical temperature in the Ising system.
The value $1/\delta$ (Ising) is $1/15$ and $1/4.8$ for the two and three-dimensional cases, and $1/3$ in the mean field theory, which becomes exact in four and higher dimensions.\cite{LC-tri-1}
The exponent determined in our analysis does not agree with any of these values.
Our result is unexpected and interesting, since criticality in thermodynamics is well described by the Ising UC.
In the Mott transition, double occupancy $\langle n_{i\uparrow}n_{i\downarrow }\rangle $ plays the role of order parameter, and the DMFT calculation shows that its dependence on
$U$ and $T$ is well fitted by the scaling function with the mean filed exponents $\beta=1/2$ and $1/\delta=1/3$.
Our result is also different from the one by Papanikolaou~{\it et al.}\cite{MTC-kapptheo-2}
Therefore, the present result implies the breakdown of the analogy with thermodynamic criticality and we need a new description for transport properties beyond the one based on free energy,
since they are nonequilibrium phenomena.

One needs some care in the scaling analysis.
We first tried the above scaling analysis with data points a little more apart from the critical point and the obtained exponents were distinct between the two sides,
$1/\delta_{-}=0.27\pm0.03$ and $1/\delta_{+}=0.11\pm0.03$.
We get to obtain $\delta_{-}$$\sim$$\delta_{+}$ after a few more data points are added closer to the critical point.
This indicates that the true critical region is rather narrow and subleading terms are not negligible outside it.
The critical region is $-0.12 \le U-U^* \le 0.11$, which is determined such that the subleading term contributes less than 2 \% there.

It is also important to note that we obtain the same exponent on the insulating side only when $D$ is chosen as a scaling variable.
We performed the same analysis for dc-conductivity $\sigma_{0}$.
Its suppression on the insulating side is more prominent than for $D$, and the obtained exponents are clearly distinct,
$(1/\delta_{-}, 1/\delta_{+})=(0.16\pm0.04, 0.07\pm0.02)$, but the value on the metallic side is close to the one determined from $D$.
This robustness is also consistent with the fact that another scaling with using only $D_{0}$, $D_{\rm 0}=A(U^*-U)^{1/\delta}$, also leads to a 
similar value $1/\delta =0.13\pm0.03$.
Scaling analysis of experimental data was performed only for the metallic side,\cite{MTC-exp-V,MTC-exp-kapp} and it is useful to
check by experiments whether the insulating side has the same exponent.

Finally, let us briefly explain our algorithm of including the vertex corrections\cite{OC-DCA} in CDMFT.
Based on the Kubo formula, $\sigma(\omega)$ is defined as
$\sigma(\omega)=-i{\rm \lim_{\mathbf q \rm \rightarrow 0}}[\chi_{\mathbf q}^J(\omega)-\chi_{\mathbf q}^J(0)]/\omega$.
Here, the current-current correlation function is described in Matsubara space as, 
\begin{eqnarray}
\chi_{\mathbf q=0}^J(i\omega_n)&=&v_{\mathbf k}^2\chi_{\mathbf k}^{0,\sigma}(i\omega_n)\nonumber \\
&+&v_{\mathbf k}v_{\mathbf k'}\chi_{\mathbf k}^{0,\sigma}(i\omega_n)\Gamma_{\mathbf k \mathbf k'}^{\sigma \sigma'}(i\omega_n)
\chi_{\mathbf k'}^{0,\sigma'}(i\omega_n),
\label{eq:OC-1}
\end{eqnarray}
where $\chi_{\mathbf k}^{0,\sigma}(i\omega_n)=-\frac{T}{N}G_{\mathbf k}^{\sigma}(i\epsilon_{l})G_{\mathbf k}^{\sigma}(i\epsilon_{l}+i\omega_n)$
and $G_{\mathbf k}^{\sigma}(i\epsilon_{l})$ is the Green's function of electron with wave vector $\mathbf k$ and spin $\sigma$.
The above equation is exact except that the dependence on $\epsilon_{l}$ and $\epsilon_{l'}$ has been averaged over in the full vertex 
$\Gamma_{\mathbf k \mathbf k'}^{\sigma \sigma'}$.
We use the convention of summing over repeated indices.
$N$ is total number of sites and $v_{\mathbf k}$ is the $x$-component of velocity $v_{\mathbf k}=2t(\rm sin\it{k_{x}}+\rm sin\frac{\it{k_{x}}}{\rm2}\rm cos\frac{\sqrt{\rm3}\it{k_{y}}}{\rm 2})$.
Spin susceptibility has been calculated including vertex corrections already in single-site DMFT approaches,\cite{SC-DMFT}
while vertex corrections in current correlations can be included only in the cluster generalization.
To obtain the full vertex $\Gamma_{\mathbf k \mathbf k'}^{\sigma \sigma'}(i\omega_n)$, we first calculate two-electron Green's functions in the cluster
$\chi_{\alpha \beta \gamma \delta}^{\sigma \sigma'}(\tau)=\langle c_{\alpha\sigma}^\dagger(\tau) c_{ \beta\sigma}(\tau)
c_{\gamma \sigma'}^\dagger(0) c_{\delta \sigma'}(0)\rangle $
by CTQMC, where $\alpha$-$\delta$ denote sites in the cluster.
From this, we obtain the irreducible vertex in the cluster by solving the Bethe-Salpeter equation 
$\chi_{\alpha \beta \gamma \delta}^{\sigma \sigma'}(i\omega_n)=\chi_{\alpha \beta \gamma \delta}^{0,\sigma \sigma'}(i\omega_n)
+\chi_{\alpha \beta \alpha' \beta'}^{0,\sigma \sigma''}(i\omega_n)
I_{ \alpha' \beta' \gamma' \delta'}^{\sigma'' \sigma'''}(i\omega_n)\chi_{\gamma' \delta' \gamma \delta}^{\sigma''' \sigma'}(i\omega_n)$,
where $\chi_{\alpha \beta \gamma \delta}^{0,\sigma \sigma'}(i\omega_n)$ denotes
the free part calculated from $G_{\mathbf k}^{\sigma}(i\epsilon_{l})$.
We then calculate the lattice irreducible vertex $I_{\mathbf k,\mathbf k'}^{\sigma \sigma'}(i\omega_n)=
I_{\alpha \beta \gamma \delta}^{\sigma \sigma'}(i\omega_n)e^{i\mathbf k(\mathbf r_{\alpha}-\mathbf r_{\beta})
+i\mathbf k'(\mathbf r_{\gamma}-\mathbf r_{\delta})}$, and obtain
$\Gamma_{\mathbf k \mathbf k'}^{\sigma \sigma'}(i\omega_n)$ via the Bethe-Salpeter equation,
$\Gamma_{\mathbf k \mathbf k'}^{\sigma \sigma'}(i\omega_n)=I_{\mathbf k \mathbf k'}^{\sigma \sigma'}(i\omega_n)
+I_{\mathbf k \mathbf k''}^{\sigma \sigma''}(i\omega_n)
\chi_{\mathbf k''}^{0,\sigma''}(i\omega_n)
\Gamma_{\mathbf k'' \mathbf k'}^{\sigma'' \sigma'}(i\omega_n)$.
We finally calculate $\sigma(\omega)$ by means of analytic continuation
$i\omega_{n}\rightarrow \omega+i 0 $ by using the maximum entropy algorithm.\cite{MEM} 
\begin{figure}
\centering
\centerline{\includegraphics[height=2in]{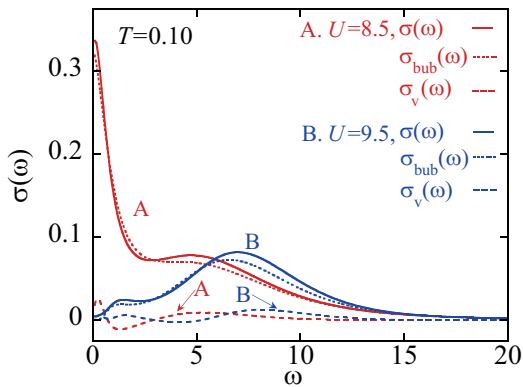}}
\caption{(Color online) Effects of vertex corrections on optical conductivity $\sigma(\omega)$.
}  
\vspace{-0.5cm}
\label{fig:tau1}
\end{figure}

Figure~\ref{fig:tau1}~presents the effects of vertex corrections;~$\sigma_{\rm bub}(\omega)$ is the results
without vertex corrections obtained from the first term in Eq.~(\ref{eq:OC-1}),
and the contribution of the vertex corrections is $\sigma_{\rm v}(\omega)=\sigma(\omega)-\sigma_{\rm bub}(\omega)$.
The data for $U=8.5$ are typical result on the metallic side.
The vertex correction changes the Drude peak into more coherent; a larger weight and a narrower width.
On the insulating side $U=9.5$, an IG peak is enhanced and the Hubbard peak slightly shifts toward a higher $\omega$
by the vertex correction.
On both sides, the total spectrum shifts toward a lower $\omega$.
For the scaling analysis of $D$, the results determined by $\sigma_{\rm bub}(\omega)$ are $(U^*, D^*)=(9.3997\pm10^{-4}, 0.09\pm0.01)$
and $(1/\delta_{-}, 1/\delta_{+})=(0.16\pm0.03, 0.12\pm0.02)$.
Detailed analysis shows that subleading terms have larger contribution compared with the case including vertex corrections.
Consequently, the critical regions are narrower, particularly on the insulating side: $-0.10 \le U-U^* \le 0.02$.
This results in larger errors in the obtained value of the exponent $1/\delta_{\pm}$.
Introducing local vertex correction partially recovers the otherwise violated charge conservation in DMFT,
and enhances coherence in charge transport.
Our results suggest that the vertex corrections also suppress subdominant modes in the critical behavior of transport and the criticality
becomes more apparent.

In this paper, we have studied optical conductivity $\sigma(\omega)$ near the Mott critical end point in the triangular-lattice Hubbard model by CDMFT
combined with a new algorithm for including vertex corrections.
We have found that $\sigma(\omega)$ has an ``ingap" peak within the Hubbard gap on the insulating side and that it continuously evolves into 
Drude peak on the metallic side.
We examined the possibility of criticality in $\sigma(\omega)$ and demonstrated that the weight of Drude and ingap peak exhibits a power-law singularity at the Mott critical point.
Although our calculations are a dynamical mean-field approximation, its critical exponent differs from the putative mean-field exponent and also from that of Ising universality class
in any dimension.
This indicates a nonequilibrium nature of transport criticality.

The authors are grateful to T. Ohashi and H. Kusunose for helpful discussions. 
The present work is supported by MEXT Grant-in-Aid for Scientific Research on Priority Areas ``Novel States of Matter Induced by Frustration" (No.~19052003),
and by Next Generation Supercomputing Project, Nanoscience Program, MEXT, Japan.
Numerical computation was performed with facilities at Supercomputer Center in Institute for Solid
State Physics and Information Technology Center,
University of Tokyo.

\end{document}